\documentclass[reprint,amsmath,amssymb,aps,prb,twocolumn,showpacs,superscriptaddress]{revtex4-1}

\usepackage{graphicx}
\begin{document}

\title{3D nature of ZrTe$_5$ band structure measured by high-momentum-resolution photoemission spectroscopy}

\author{H. Xiong}
\affiliation{Geballe Laboratory for Advanced Materials, Departments of Physics and Applied Physics, Stanford University, Stanford, CA 94305, USA}
\affiliation{Stanford Institute for Materials and Energy Sciences, SLAC National Accelerator Laboratory, 2575 Sand Hill Road, Menlo Park, CA 94025, USA}

\author{J. A. Sobota}
\affiliation{Geballe Laboratory for Advanced Materials, Departments of Physics and Applied Physics, Stanford University, Stanford, CA 94305, USA}
\affiliation{Stanford Institute for Materials and Energy Sciences, SLAC National Accelerator Laboratory, 2575 Sand Hill Road, Menlo Park, CA 94025, USA}
\affiliation{Advanced Light Source, Lawrence Berkeley National Laboratory, Berkeley, California 94720, USA}

\author{S.-L. Yang}
\affiliation{Geballe Laboratory for Advanced Materials, Departments of Physics and Applied Physics, Stanford University, Stanford, CA 94305, USA}
\affiliation{Stanford Institute for Materials and Energy Sciences, SLAC National Accelerator Laboratory, 2575 Sand Hill Road, Menlo Park, CA 94025, USA}

\author{H. Soifer}
\affiliation{Stanford Institute for Materials and Energy Sciences, SLAC National Accelerator Laboratory, 2575 Sand Hill Road, Menlo Park, CA 94025, USA}

\author{A. Gauthier}
\affiliation{Geballe Laboratory for Advanced Materials, Departments of Physics and Applied Physics, Stanford University, Stanford, CA 94305, USA}
\affiliation{Stanford Institute for Materials and Energy Sciences, SLAC National Accelerator Laboratory, 2575 Sand Hill Road, Menlo Park, CA 94025, USA}

\author {M.-H. Lu}
\affiliation{National Laboratory of Solid State Microstructures, Department of Materials Science and Engineering, Nanjing University, Nanjing, Jiangsu 210093, China}

\author{Y.-Y. Lv}
\affiliation{National Laboratory of Solid State Microstructures, Department of Materials Science and Engineering, Nanjing University, Nanjing, Jiangsu 210093, China}

\author{S.-H. Yao}
\affiliation{National Laboratory of Solid State Microstructures, Department of Materials Science and Engineering, Nanjing University, Nanjing, Jiangsu 210093, China}

\author{D. Lu}
\affiliation{Stanford Synchrotron Radiation Lightsource, SLAC National Accelerator Laboratory, 2575 Sand Hill Road, Menlo Park, California 94025, USA.}

\author{M. Hashimoto}
\affiliation{Stanford Synchrotron Radiation Lightsource, SLAC National Accelerator Laboratory, 2575 Sand Hill Road, Menlo Park, California 94025, USA.}

\author{P. S. Kirchmann}
\affiliation{Stanford Institute for Materials and Energy Sciences, SLAC National Accelerator Laboratory, 2575 Sand Hill Road, Menlo Park, CA 94025, USA}

\author{Y.-F. Chen}
\affiliation{National Laboratory of Solid State Microstructures, Department of Materials Science and Engineering, Nanjing University, Nanjing, Jiangsu 210093, China}
\affiliation{Collaborative Innovation Center of Advanced Microstructures, Nanjing University, Nanjing, Jiangsu 210093, China}

\author{Z.-X. Shen}
\email{zxshen@stanford.edu}
\affiliation{Geballe Laboratory for Advanced Materials, Departments of Physics and Applied Physics, Stanford University, Stanford, CA 94305, USA}
\affiliation{Stanford Institute for Materials and Energy Sciences, SLAC National Accelerator Laboratory, 2575 Sand Hill Road, Menlo Park, CA 94025, USA}

\date{\today}

\begin{abstract}
We have performed a systematic high-momentum-resolution photoemission study on ZrTe$_5$ using $6$~eV photon energy. We have measured the band structure near the $\Gamma$ point, and quantified the gap between the conduction and valence band as $18 \leq \Delta \leq 29$~meV. We have also observed photon-energy-dependent behavior attributed to final-state effects and the 3D nature of the material's band structure. Our interpretation indicates the gap is intrinsic and reconciles discrepancies on the existence of a topological surface state reported by different studies. The existence of a gap suggests that ZrTe$_5$ is not a 3D strong topological insulator nor a 3D Dirac semimetal. Therefore, our experiment is consistent with ZrTe$_5$ being a 3D weak topological insulator.
\end{abstract}

\maketitle


\section{Introduction}

In the past decade, topological materials such as 2D and 3D topological insulators (TI) \cite{Konig2007,Xia2009,Topological2009} and 3D Dirac and Weyl semimetals \cite{Liu2014,Liu2014a,Neupane2014,Xu2015,Lv2015} have been continuously attracting the interest of the condensed matter physics community, because of their unique band structures \cite{Xia2009,Liu2014,Liu2014a,Neupane2014,Xu2015,Lv2015} and transport properties \cite{Konig2007}.
Known for its large thermo-power and resistivity anomaly \cite{Jones1982,Skelton1982,Tritt1999,McIlroy2004}, and for the recent discovery of a superconducting phase under high pressure \cite{Zhou2016}, ZrTe$_5$ is a new promising platform to study topological phase transitions. 

Density functional theory (DFT) calculations have predicted ZrTe$_5$ to be a 3D strong TI with the experimentally-determined inter-layer lattice parameter, and to be a 3D weak TI with the lattice parameter $2\%$ enlarged \cite{Weng2014}. 
In the weak TI scenario, the material exhibits an energy gap between the conduction band (CB) and valence band (VB). In the strong TI scenario, the CB and VB are inverted near $\Gamma$, which leads to the formation of a gapless topological surface state (TSS).
Between the two cases, there is a lattice parameter where the 3D weak TI to strong TI phase transition happens, and the bulk CB touches the VB to form a bulk Dirac cone, which is the 3D Dirac semimetal scenario \cite{Manzoni2016}.
The parameter sensitivity within such a small range makes it challenging to experimentally establish whether ZrTe$_5$ is 3D weak or strong TI. These scenarios are distinguished by measuring the band structure of ZrTe$_5$ near the $\Gamma$ point to determine whether there is a finite band gap and whether there is a TSS. 

\begin{figure*}
\resizebox{0.85\linewidth}{!}{\includegraphics{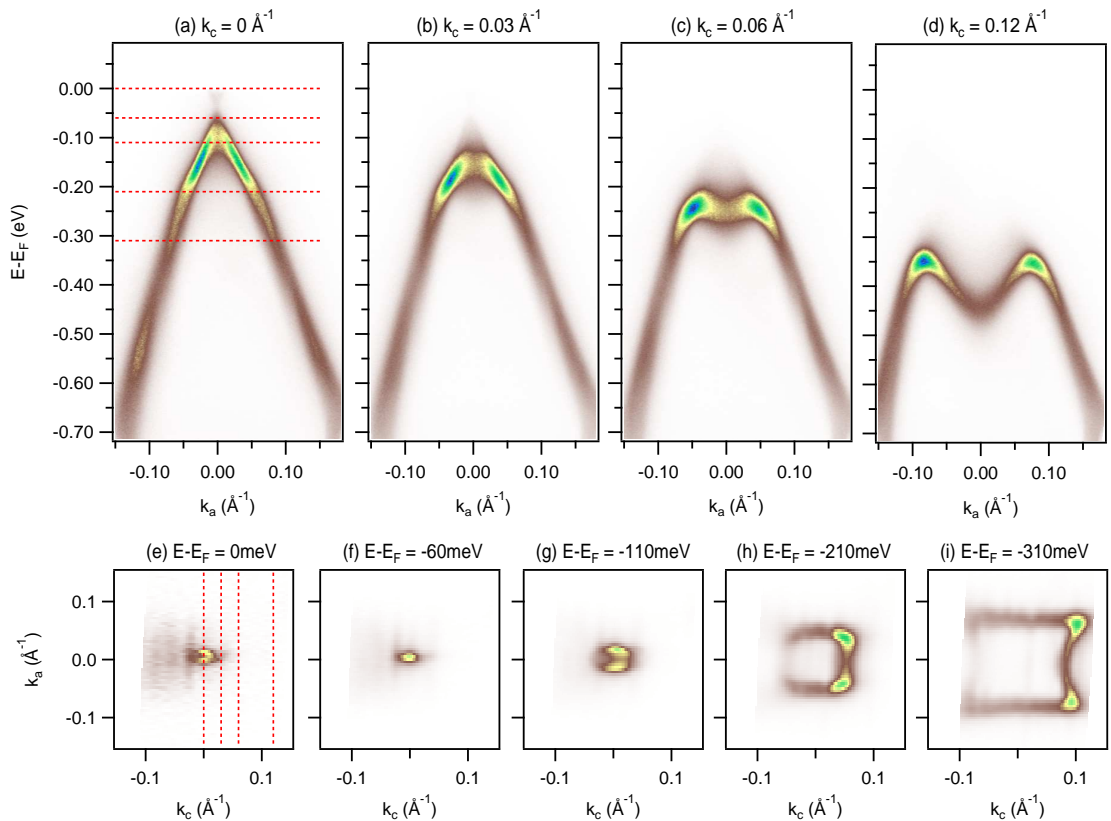}}
\caption{(a-d) Band dispersion measured along the $k_a$-axis in momentum space; the cuts are taken at $k_c = 0, 0.03, 0.06, 0.12$ \AA$^{-1}$ (red dashed lines in (e)). (e-i) Constant energy contour mapping at different energy cuts: $E-E_{\textrm{F}} = 0, -0.06, -0.11, -0.21, -0.31$ eV (red dashed lines in (a)). These measurements were performed using $5.90$ eV photon energy at $20$~K. }
\label{Band}
\end{figure*}

The discovered chiral magnetic effect in ZrTe$_5$ \cite{Li2016} and some photon-energy-dependent ARPES measurements \cite{Li2016,Shen2016} suggested a 3D Dirac semimetal band structure. Magneto-optical \cite{Chen2015,Chen2015a} and transport \cite{Zheng2016} measurements also suggested the possibility of ZrTe$_5$ being a 3D Dirac semimetal. On the other hand, a subsequent ARPES study resolved the CB by dividing the spectrum by the Fermi-Dirac function \cite{Wu2016}; combined with complementary STM results, the study reported a $100$~meV gap and concluded that ZrTe$_5$ is a 3D weak TI. Another STM measurement reported an $80$~meV gap \cite{Li2016a}. An ultrafast 2-photon-photoemission measurement directly measured the unoccupied CB, and estimated a $50$~meV upper bound to the gap size \cite{Manzoni2015}. An ARPES measurement performed at $2$~K estimates the gap to be $40$~meV \cite{Zhang2016}. Based upon these different results, a more definitive measurement of the gap size is needed to understand the topological categorization of ZrTe$_5$.

An important related aspect is the existence of a TSS. One recent ARPES study reported split-band structure \cite{Manzoni2016a}, and combined with photon-energy dependence \cite{Manzoni2016} concluded that ZrTe$_5$ is a 3D strong TI with gapless TSS. On the other hand, another study reported no TSS, despite also performing photon-energy dependent experiments \cite{Moreschini2016}. Hence, a unified interpretation is needed to reconcile these discrepancies. 

In this paper, we have measured the CB and VB of ZrTe$_{5}$ near $\Gamma$, and quantified a gap $18 \leq \Delta \leq 29$~meV, using a high-momentum-resolution laser-ARPES setup with $6$~eV photon energy. The gap size is smaller than that reported by former ARPES and STM studies \cite{Wu2016,Li2016a,Manzoni2015,Zhang2016}. As ZrTe$_{5}$'s band structure exhibits a binding energy shift as a function of temperature \cite{Manzoni2015,McIlroy2004,Zhang2016}, we have performed a thorough temperature-dependent measurement, to determine a suitable temperature for gap analysis. At the same time, we have discovered that the binding energy shifts with a slope nearly identical to that of the work function, which we attribute to a temperature-dependent doping change. 
Finally, we examine the spectral difference between ARPES measurements performed at different photon energies; we attribute this difference to final-state effects and the 3D nature of ZrTe$_5$'s band structure. By doing so, we confirm the gap between the CB and VB is not due to some specific out-of-plane momentum $k_b $, and conclude that there is no TSS at $\Gamma$. The final-state interpretation reconciles the discrepancies of previous studies regarding the existence of TSS. Our result is consistent with ZrTe$_5$ being a 3D weak TI.


\section{Methods}

\begin{figure}
\resizebox{1.05\linewidth}{!}{\includegraphics{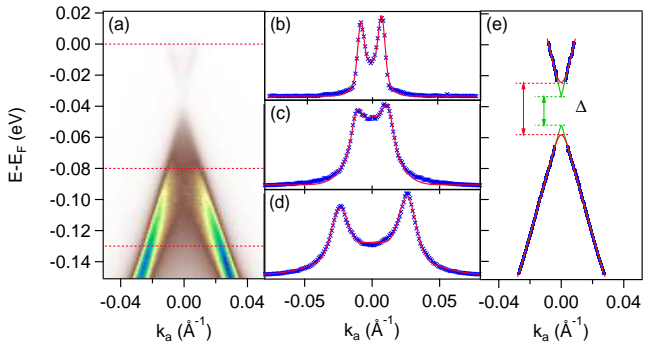}}
\caption{(a) The spectrum containing both CB and VB with a gap in between. (b-d) Momentum distribution curves (MDCs) at corresponding binding energies in (a). (e) Within a $0.05$~\AA$^{-1}$ momentum range, both CB and VB dispersions are extracted from MDC fitting (blue dots). We use two different models to fit the band structure. A simple linear extrapolation fitting gives a gap size of $18\pm2$~meV (green lines); another more complex model (see in text) gives a gap size of $29\pm7$~meV (red curves). These measurements were performed using $5.90$~eV photon energy at $20$~K. }
\label{Gap}
\end{figure}

Polycrystalline ZrTe$_5$ samples were first prepared by the direct stoichiometric solid-state reaction of high pure Zr (5N) with Te powder (5N) in a fused silica tube sealed under vacuum pressure around $4\times10^{-6}$~Torr at about $500^{\circ}$C for 7 days. Then ZrTe$_5$ polycrystals and about $3$ mg$\cdot$cm$^{-3}$ of high purity iodine (I2) were ground and loaded into an evacuated quartz ampoule, and then placed into a double zone furnace with a temperature profile of $450\sim550^{\circ}$C to grow crystals. The millimeter-sized strip single crystals with metallic luster were successfully obtained after growth of a period over $10$ days \cite{Lv2016}. The experimental lattice constants  at $300$~K are $a=3.9943$\AA, $b=14.547$\AA, and $c=13.749$\AA. 

Samples were cleaved \textit{in situ} at a base pressure lower than $5\times10^{-11}$~Torr. ARPES measurements were carried out using Scienta R4000 electron analyzer and a tunable Ti:Sapphire oscillator with photon energy quadrupling through two stages of second harmonic generation, to output $6$~eV ultraviolet light. ARPES spectra were acquired over a photon energy range from $5.6$~eV to $6.0$~eV. The energy, angular, and momentum resolution for this setup are $8$~meV, $0.3^{\circ}$, and $0.004$~\AA$^{-1}$ respectively. Preliminary characterization was performed at the Stanford Synchrotron Radiation Lightsource. 


\section{Results}
We show the band dispersion and energy contour maps of ZrTe$_5$ in Fig.~\ref{Band}, which is measured at $20$~K. Fig.~\ref{Band}(a-d) show the band structure measured along the $k_a$-axis in momentum space, at different $k_c = 0, 0.03, 0.06, 0.12$~\AA$^{-1}$. The cut (a) taken through $\Gamma$ reveals that the VB is $\Lambda$-shaped with its top $\sim50$~meV below the Fermi energy ($E_{\textrm{F}}$), and the CB is V-shaped. Away from $\Gamma$, the CB disperses above $E_{\textrm{F}}$ and the VB disperses to lower energies, gradually evolving into an M-shape, as shown in Fig.~\ref{Band}(b-d). Fig.~\ref{Band}(e-i) show the constant energy contour mappings at different energies: $E-E_{\textrm{F}} = 0, -0.06, -0.11, -0.21, -0.31$~eV. The small elliptical contour corresponding to the CB in Fig.~\ref{Band}(e) gradually evolves to a single point between the CB and VB in Fig.~\ref{Band}(f), and then to curved rectangle corresponding to the VB in Fig.~\ref{Band}(g-i). The band dispersion and energy contour maps together reveal that both the CB and VB are cone-like near $\Gamma$.

In Fig.~\ref{Gap}, we quantify the band dispersion by fitting the momentum distribution curves (MDCs) in a small range around $\Gamma$.
We take MDCs at different binding energies and fit each MDC with a function composed of two Gaussian-form peaks plus a Gaussian background. We choose three binding energies in Fig.~\ref{Gap}(a)(red dashed lines) and plot the MDCs in Fig.~\ref{Gap}(b-d)(blue crosses). We can see in Fig.~\ref{Gap}(b-d) that the fitting function adequately fits the MDCs (red solid curves). The fitted peak positions are plotted in Fig.~\ref{Gap}(e), capturing both CB and VB; a $0.004$~\AA$^{-1}$ linewidth and $15$~meV lifetime at $E_{\textrm{F}}$ is given by the fitting. 
Near the top of the VB and the bottom of the CB where peaks are overlapping, there is a large uncertainty to fit the peak positions, so we avoid this energy range in the analysis.

\begin{figure*}
\resizebox{0.75\linewidth}{!}{\includegraphics{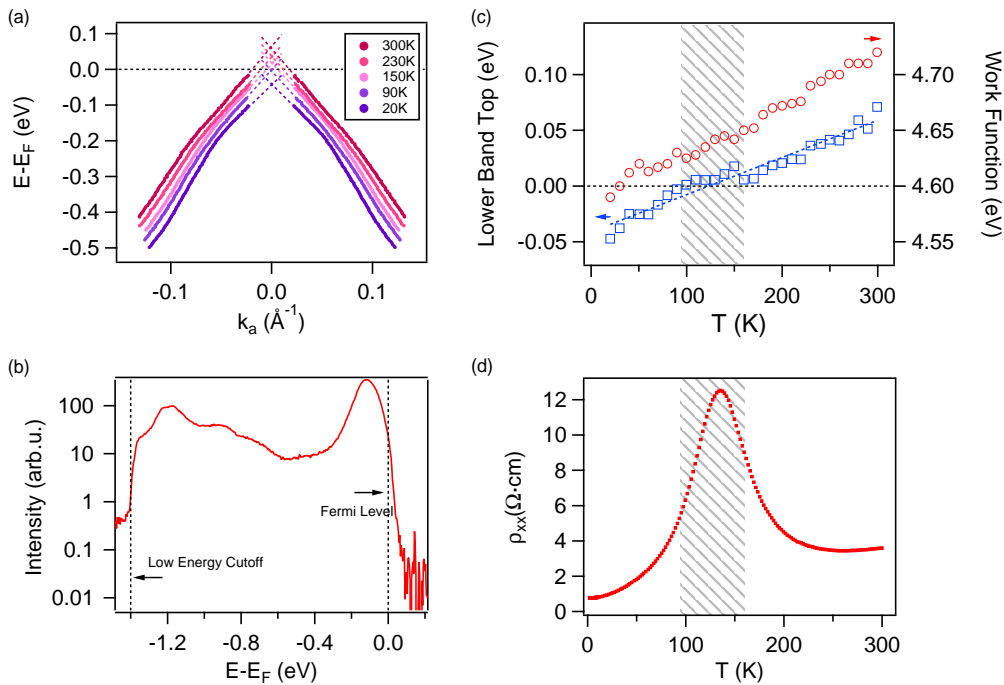}}
\caption{(a) There is a monotonic binding energy shift when the temperature increases from $20$~K to $300$~K. The band top is extrapolated (intersects of the dashed lines) and shifts from below to above $E_{\textrm{F}}$. (b) The momentum-integrated energy distribution curve is used to extract the low energy cutoff, which is then used to calculate the work function. (c) The energy of the VB top at different temperatures is plotted in blue squares; the work function at different temperatures is plotted in red circles. These measurements were performed using $6.02$~eV photon energy. (d) The transport measurement shows the resistivity along the $a$-axis peaks at $140$~K, falling into the shaded region in (c). }
\label{Temperature}
\end{figure*}

In order to quantify the gap between the CB and VB, we use two different models to fit the band dispersion. 
First, we linearly fit the bands and obtain the VB and CB velocities as $3.3$~eV$\cdot$\AA~and $4.1$~eV$\cdot$\AA, respectively. By extrapolating both the CB and VB we obtain a gap of $18$~meV (Fig.~\ref{Gap}(e)). Since the linear extrapolation neglects the fact that both the VB top and the CB bottom should have a finite curvature, this fitting model underestimates the gap size. 
To capture the finite curvature in the vicinity of the gap, as well as the difference in band slopes, we considered a more complex model to fit both the CB and VB: 
$f(k) = A(k^2 + k_{0}^{2}) + B \pm \sqrt[]{4A^2k^2k_{0}^{2}+\dfrac{\Delta^2}{4}}$
, where $A$ is a coefficient related to the curvature of the band, inversely proportional to electron effective mass; $B$ is an offset that determines $E_{\textrm{F}}$; $\Delta$ is the gap between the CB and VB; and $k_0$ is a momentum offset considerably larger than the momentum range where data are fitted. 
The fitting is shown in a solid red line and gives a gap size of $29$~meV (Fig.~\ref{Gap}(e)). The fitting has a larger deviation from the data near the gap, which yields an overestimation of the gap size. 
Considering the results from both models, we conclude that ZrTe$_5$ possesses a gap $18 \leq \Delta \leq 29$~meV between the CB and VB.

We have also conducted a thorough temperature-dependent measurement increasing from $20$~K to $300$~K with $10$~K spacing, to study the binding energy shift as a function of temperature. This analysis shows that the gap is best studied at low temperature, where the CB is below $E_{\textrm{F}}$. Fig.~\ref{Temperature}(a) plots MDC-fitted band structures at five different temperatures; the band shifts monotonically to higher energy with increasing temperature. For each temperature, the VB top was extrapolated based on a linear fitting, as shown in Fig.~\ref{Temperature}(a)(dashed lines) and (c)(blue squares). The shading in Fig.~\ref{Temperature}(c,d) highlights the temperature region where the VB top crosses $E_{\textrm{F}}$. In order to understand the origin of this temperature-dependent binding energy shift, we also measure the corresponding work function, as shown in Fig.~\ref{Temperature}(b). The momentum-integrated energy distribution curve (EDC) shows two intensity drops at both high and low energy. The high energy drop corresponds to $E_{\textrm{F}}$; the more abrupt intensity drop at low energy is the low energy cut-off, below which photo-excited electrons are not able to overcome the material's work function to be emitted from the sample. Thus, the corresponding energy $E_{\textrm{low}}$ is extracted and used to calculate work function: $W = E_{\textrm{low}} - E_{\textrm{F}} + h\nu$, where $h\nu$ is the photon energy. The work function as a function of temperature is plotted in Fig.~\ref{Temperature}(c) (red circles). As we can see, the work function has nearly the same slope with respect to temperature as the binding energy shift.
For comparison, Fig.~\ref{Temperature}(d) shows the resistivity of ZrTe$_5$ along the crystallographic $a$-axis as a function of temperature. The resistivity peaks at $140$~K, which is known as the resistivity anomaly \cite{Jones1982,Skelton1982,Tritt1999,McIlroy2004}.


\section{Discussion}
\begin{figure*}
\resizebox{\linewidth}{!}{\includegraphics{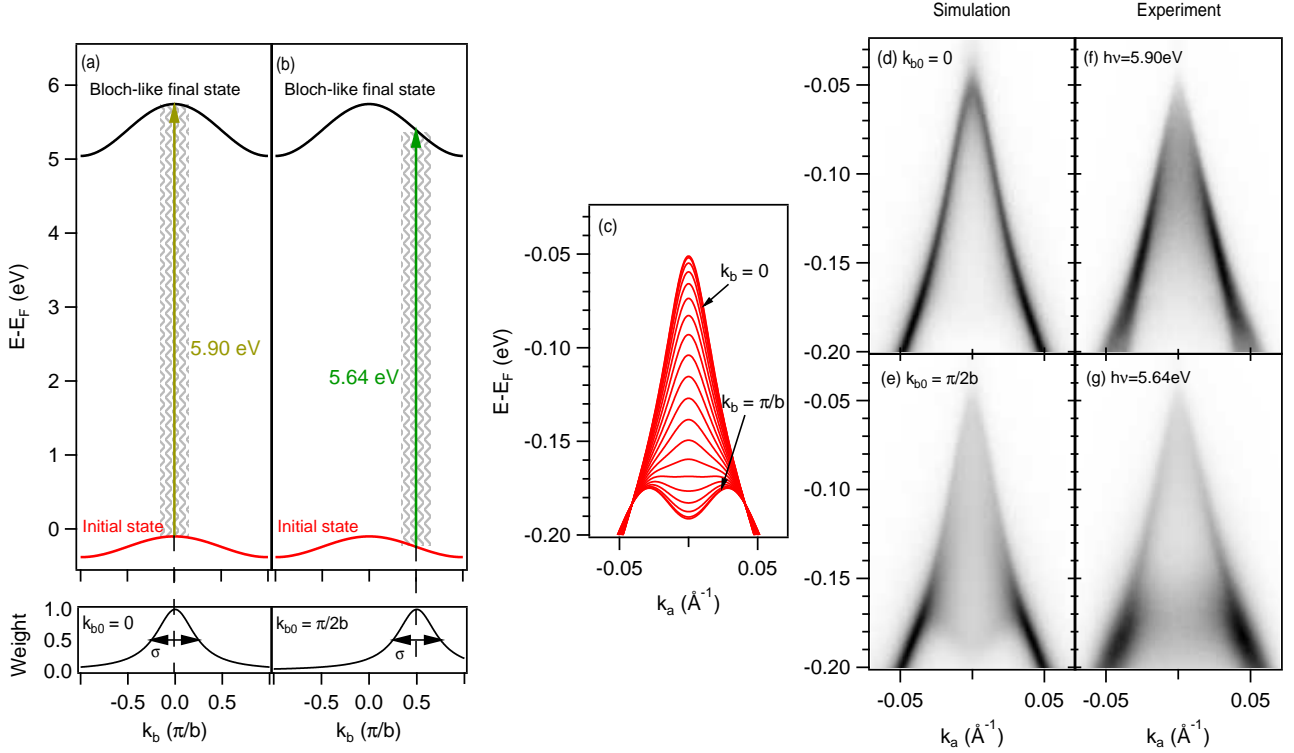}}
\caption{(a,b) Cartoons showing that photons with different energies photoemit electrons with different initial-state momenta along the $k_b$-axis. Both the initial- and final-state dispersions are plotted along $k_b$. The bottom row shows that for each photon energy, instead of a single $k_b$, a Lorentzian distribution centered at $k_{b0}$ with width $\sigma$ contributes to the photoemission process. Due to the Bloch-like final-state, even a small change of photon energy shifts $k_{b0}$ by a significant portion of the Brillouin zone. (c) Band dispersion at different $k_b$ constructed to phenomenologically reproduce the experiment. (d,e) Simulated spectra with $\sigma=0.3\pi/b$ at $k_{b0}=0,\pi/2b$, respectively. The parameters are chosen to reproduce the experiments. (f,g) ARPES spectra measured at $h\nu = 5.90$~eV and $5.64$~eV, respectively. }
\label{Dependence}
\end{figure*}

\subsection{Gap Size, Linewidth, \& Binding Energy Shift}
Currently the most interesting question regarding ZrTe$_5$ is whether it is a 3D Dirac semimetal \cite{Li2016,Chen2015,Chen2015a}, a 3D weak TI \cite{Wu2016,Li2016a}, or a 3D strong TI \cite{Manzoni2016}. The most essential evidence to distinguish these scenarios is to determine whether it has a gapless Dirac cone near $\Gamma$, or a gapped state, and to determine the existence of a TSS. 
Previous ARPES measurements reported gaps ranging from $0$ to $100$~meV \cite{Wu2016,Li2016a,Shen2016,Manzoni2015,Zhang2016}. In the case of zero gap \cite{Li2016a,Shen2016}, the CB was not resolved, making it difficult to have a conclusive measurement concerning the existence of a gap. As for the remaining works with gaps ranging from $\sim100$~meV \cite{Wu2016,Li2016a} to below $50$~meV \cite{Manzoni2015,Zhang2016}, the variability could be attributed to measurements taken at $k_b \neq 0$, differences in resolution, or subtle differences in lattice parameter \cite{Weng2014}. We have measured a gap $18\leq \Delta \leq 29$~meV, smaller than previously reported, setting a new benchmark in the studies of ZrTe$_5$ with gapped band structure. 

Another interesting note is that the $0.004$~\AA$^{-1}$ linewidths we extract from the MDC fitting are among the smallest linewidths ever measured by ARPES, even compared to what have been reported on the TSS of TIs \cite{Pan2012,Kondo2013}. 
Therefore, the high-momentum-resolution ARPES measurement enables us not only to quantify a smaller gap, but also to resolve a small linewidth, which might indicate weak quasi-particle scattering in ZrTe$_5$ \cite{Pan2012,Kondo2013}.

\subsection{Binding Energy Shift \& Resistivity Anomaly}
Our temperature-dependent measurement goes down to temperatures low enough to directly see the CB, which is critical for quantifying a gap. Moreover, the binding energy shift as a function of temperature shows nearly identical slope as that of the work function change, implying a temperature-dependent doping-level change, rather than a change in the crystal structure \cite{Skelton1982}. 

Interestingly, G. Manzoni \textit{et al.} \cite{Manzoni2016a} show a non-monotonic binding energy shift with a turning temperature at $150$~K; Y. Zhang \textit{et al.} \cite{Zhang2016} report a similar monotonic binding energy shift as us from 300K to 2K; while L. Moreschini \textit{et al.} \cite{Moreschini2016} shows an opposite shifting direction. From these scattered results, we speculate that the temperature-dependence of the binding energy shift is highly dependent on the sample growth condition and/or surface condition. However, regardless of the mechanism of the temperature-dependent doping level, we believe it is directly correlated to the work function. 

The resistivity peaks at $140$~K, falling into the shaded region in Fig.~\ref{Temperature}(c), indicating a direct correspondence between the photoemission result and transport measurement: when the top of VB crosses $E_{\textrm{F}}$, the density of states at $E_{\textrm{F}}$ is minimal, which is consistent with a maximum in resistivity \cite{McIlroy2004,Manzoni2015,Zhang2016}.

\subsection{Final-state Effect \& 3D Nature of Band Structure}
\begin{figure*}
\resizebox{0.62\linewidth}{!}{\includegraphics{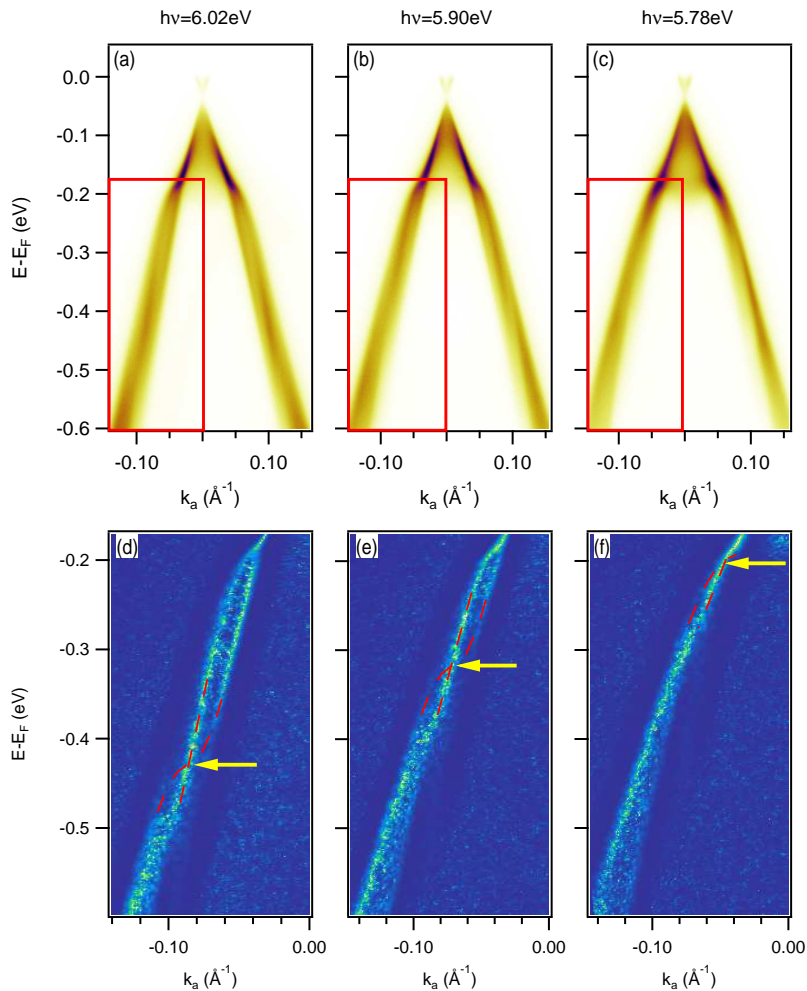}}
\caption{(a-c) ARPES spectra measured using photon energies $h\nu = 6.02, 5.90, 5.78$~eV. (d-f) Spectra processed by the minimum gradient method \cite{He2016}, each corresponding to the red rectangular region in (a-c). The red dashed curves are guides-to-the-eye following the intensity peaks, and the yellow arrow shows where the ``momentum switching" occurs.}
\label{Twist}
\end{figure*}

In order to make sure the gap is not due to a specific out-of-plane momentum $k_b$, we have investigated the band structure using different photon energies $5.90$~eV and $5.64$~eV (Fig.~\ref{Dependence}(f,g)): the gap sizes using the linear extrapolation method are $18$~meV and $19$~meV for $5.90$~eV and $5.64$~eV, respectively, therefore consistent across different photon energies. However, we observe significant difference between the two spectra measured: the $5.90$~eV spectrum shows a sharp $\Lambda$-shaped band; the $5.64$~eV spectrum shows the $\Lambda$-shape band with reduced intensity, but at the same time, there is an M-shaped band $\sim140$~meV below the top of the VB.

We interpret the spectral difference as a final-state effect, and due to the 3D nature of ZrTe$_5$'s band structure \cite{Moreschini2016}. For the following discussion, we adopt the more general convention for ARPES, using $k_{\vert\vert}$ and $k_{\perp}$ to describe the in-plane and out-of-plane momentum directions. In general, the spectrum measured by photoemission spectroscopy can be approximated as \cite{Lindroos1996}: 
\begin{equation}
I(k_{\vert\vert},\omega)\propto \int |M_{if}|^2 A_f(k_{\vert\vert},k_{\perp}, \omega + h\nu)
 A_i(k_{\vert\vert},k_{\perp}, \omega) dk_{\perp}
\end{equation}
where 
$|M_{if}|$ is the matrix element between initial state and final state, which we assume is constant in the following discussion \cite{Lindroos1996}; 
$A_i$($A_f$) is the initial state (final state) 
spectral weight; $k_{\vert\vert}$ and $k_{\perp}$ denote the in-plane and out-of-plane momentum; and $h\nu$ and $\omega$ are the photon energy and the electron initial state energy, respectively. The final-state spectral weight is represented by a Lorentzian function:
\begin{equation}
A_f(k_{\vert\vert},k_{\perp}, \omega + h\nu) \propto \dfrac{\Sigma''}{(\omega + h\nu - E_{k_{\vert\vert}, k_{\perp}} - \Sigma')^2 + (\Sigma'')^2}
\end{equation}
where $E_{k_{\vert\vert}, k_{\perp}}$ is the dispersion of final state, and $\Sigma'(\Sigma'')$ is the real (imaginary) part of the self-energy.

If we define $E_{k_{\vert\vert}, k_{0}} = \omega + h\nu - \Sigma'$ and make a local linear approximation to the final state dispersion with respect to $k_{\perp}=k_0$, we can express: $E_{k_{\vert\vert}, k_{\perp}} = E_{k_{\vert\vert}, k_{0}} + v_{\perp}(k_{\perp}-k_{0})$, where $v_{\perp} = (\partial E_{k_{\vert\vert}, k_{\perp}}/\partial k_{\perp}) |_{k_{0}}$, so the final-state spectral function could be expressed as:
\begin{equation}
A_f(k_{\vert\vert},k_{\perp}, \omega + h\nu) \propto \dfrac{\sigma}{(k_{\perp} - k_{0})^2 + \sigma^2}
\end{equation}
where $\sigma = \Sigma''/v_{\perp}$; and so we can evaluate the spectrum as: 
\begin{equation}
I(k_{\vert\vert},\omega)\propto \int \dfrac{\sigma A_i(k_{\vert\vert},k_{\perp}, \omega)}{(k_{\perp} - k_{0})^2 + \sigma^2} dk_{\perp}
\end{equation}
In this scenario, the spectrum measured at each different photon energy will correspond to a distribution of initial state $k_{\perp}$ centered at $k_{0}$ with a finite width $\sigma$. 

In addition, if the photon energy is not high enough, the emitted electrons are not completely free, but driven to unoccupied Bloch-like final states with flat dispersion\cite{Lindroos1996}. This explains the possibility that even a small change in photon energy could induce a large shift of out-of-plane momentum $k_{0}$, even a significant portion of Brillouin zone. This is completely different in the case of free-electron-like final state, where the final state dispersion has a larger slope. If we assume a free-electron final state probed with photon energy near $6$~eV and consider an inner potential $V = 16$~eV \cite{Li2016}, and work function $W = 4.6$~eV, then for a change of $\Delta h\nu = 0.26$~eV
, the shift of out-of-plane momentum would be $\Delta k_{\perp} = 0.015$~\AA$^{-1}$
. This is only $\sim 5\%$ of the Brillouin zone in ZrTe$_5$, which would not cause a significant spectral change. Therefore, the large spectral change observed experimentally is related to Bloch-like final states.

We build a toy model for ZrTe$_5$ to illustrate this interpretation, as shown in Figure 4. As the sample is cleaved along the $a$-$c$ plane, the out-of-plane momentum $k_{\perp}$ for ZrTe$_5$ is $k_b$. We have phenomenologically constructed an analytic form for the band dispersion to qualitatively reproduce our experimental spectra.
The dispersion along the $k_a$-axis at different $k_b$ is shown in Fig.~\ref{Dependence}(c); it captures the $\Lambda$-shaped band and the M-shaped band measured in Fig.~\ref{Dependence}(f,g), which are the extrema of the band dispersion along $k_b$. 
For $5.90$~eV photon energy, electrons are excited to final states that are not completely free (Fig.~\ref{Dependence}(a)); considered the Lorentzian broadening, the simulated spectrum shows a prominent $\Lambda$-shaped feature (Fig.~\ref{Dependence}(d)). For $5.64$~eV photon energy (Fig.~\ref{Dependence}(b)), considered the Lorentzian broadening and the large out-of-plane momentum shift after changing photon energy, the simulated spectrum shows both $\Lambda$-shaped and M-shaped features (Fig.~\ref{Dependence}(e)). 
The figures at the bottom of Fig.~\ref{Dependence}(a,b) qualitatively visualize the large out-of-plane momentum shift of the initial state contribution at the two photon energies. The parameters shown in the figures are chosen to reproduce the experiments. 
Comparing the simulations in Fig.~\ref{Dependence}(d,e) with the experiments in Fig.~\ref{Dependence}(f,g), we can see the model based on final-state-effect explains the data very well. 

Hence, the spectral difference between $5.64$~eV and $5.90$~eV could be sufficiently understood in terms of a final-state effect \cite{Lindroos1996}. The photon-energy-dependent spectra reflect the material's intrinsic 3D band structure, and the consistency of gap size indicates the gap between the CB and VB is representative of the true band-gap, as each spectrum contains a finite contribution from $k_b = 0$. 


Our final-state-effect interpretation may even reconcile discrepancies regarding the existence of a TSS raised by recent ARPES studies \cite{Manzoni2016a,Moreschini2016,Manzoni2016}. G. Manzoni \textit{et al.} \cite{Manzoni2016} interpreted the $\Lambda$-shaped band as a TSS due to $k_b$ independent spectral weight at a particular binding energy. However, the band dispersion with respect to $k_b$  (shown by our model in Fig. 4(a) and supported by DFT calculations \cite{Weng2014,Manzoni2016}) features a near-isosbestic point at $k_a \approx 0.04$~\AA$^{-1}$.  Therefore, even bulk bands in the vicinity of this momentum are expected to exhibit little $k_b$ dispersion, which complicates the assignment of surface bands. On the other hand, L. Moreschini \textit{et al.} \cite{Moreschini2016} clearly resolve the $\Lambda$-shaped band evolve into an M-shape as a function of $k_b$, with no indication of a surface band. This result is consistent with the interpretation of our spectra being characterized by bulk bands modulated by final-state effects.


In fact, the relevance of final-state effects is unambiguous when we look at momenta further from $\Gamma$ (Fig.~\ref{Twist}). In Fig.~\ref{Twist} we show spectra with a larger momentum and energy range taken at three different photon energies. From Fig.~\ref{Twist}(a-c) we see what appears to be multiple bands (as indicated in the red rectangular regions), similar to the splitting recently reported \cite{Manzoni2016a}. To make the multiple bands more visually noticeable, we use the minimum gradient method \cite{He2016} to process the red rectangular region in each spectrum, as we plot in Fig.~\ref{Twist}(d-f). The resulting dispersions cannot be well described as two distinct bands, as the bands ``twist" around each other (as guided by the red dashed curves), and seem to ``switch momenta" at certain binding energies (as indicated by the yellow arrows), which change roughly by the same amount as the difference in photon energy. This behavior suggests that the spectrum is modulated by the structure of final states \cite{Miller2015}, and therefore cannot be understood solely in the context of a well-defined initial state dispersion. 
More detailed calculations involving electron final-states are required to fully understand this complexity \cite{Arrala2016}.



\section{Conclusions}
 
In this paper, we report a systematic high-momentum-resolution $6$~eV-laser photoemission study on ZrTe$_5$. We have measured a clear band structure near $\Gamma$, and quantified a gap $18 \leq \Delta \leq 29$~meV between the CB and VB. The temperature-dependent study shows nearly identical slopes of binding energy shift and work function change, indicating a temperature-dependent doping-level variation instead of a structural change. 
We have also studied the spectral difference between different photon energies and attributed it to a final-state effect, which reveals the 3D nature of ZrTe$_5$'s band structure. This interpretation suggests that there is a finite band gap between the CB and VB. This leads us to conclude that ZrTe$_5$ is neither a 3D strong TI, nor a Dirac semimetal. However, our observations are consistent with it being a 3D weak TI, though we cannot verify the existence of topological edge states \cite{Wu2016,Li2016a,Zhang2016}.

It is also worth mentioning that the measured band structure seems extremely sensitive to the material's lattice parameter \cite{Weng2014}, and there are studies proposing a topological phase transition in ZrTe$_5$ by changing the inter-layer lattice parameter \cite{Manzoni2016,Zhang2016}. 
Therefore, an interesting subject of future research would be to measure the band structure as a function of a tunable applied strain \cite{Weng2014}. 


\begin{acknowledgments}

The photoemission work was supported by the U.S. Department of Energy, Office of Science, Basic Energy Sciences, Materials Sciences and Engineering Division under contract DE-AC02-76SF00515. H.X. and J.A.S. were in part supported by the Gordon and Betty Moore Foundations EPiQS Initiative through Grant GBMF4546. S.-L.Y. acknowledges support by the Stanford Graduate Fellowship. H.S. acknowledges support from the Fulbright Scholar Program. A.G. acknowledges the support from National Defense Science \& Engineering Graduate Fellowship Program. Y.-F. C. acknowledges the National Basic Research Program of China (Grants 2013CB632700) and the National Natural Science Foundation of China 51472112 and M.-H. L. also acknowledges the support of Natural Science Foundation of Jiangsu Province (BK20140019). Stanford Synchrotron Radiation Lightsource is operated by the U.S. Department of Energy, Office of Science, Office of Basic Energy Sciences.
    
\end{acknowledgments}

\bibliography{ZrTe5}

\begin{thebibliography}{33}%
\makeatletter
\providecommand \@ifxundefined [1]{%
 \@ifx{#1\undefined}
}%
\providecommand \@ifnum [1]{%
 \ifnum #1\expandafter \@firstoftwo
 \else \expandafter \@secondoftwo
 \fi
}%
\providecommand \@ifx [1]{%
 \ifx #1\expandafter \@firstoftwo
 \else \expandafter \@secondoftwo
 \fi
}%
\providecommand \natexlab [1]{#1}%
\providecommand \enquote  [1]{``#1''}%
\providecommand \bibnamefont  [1]{#1}%
\providecommand \bibfnamefont [1]{#1}%
\providecommand \citenamefont [1]{#1}%
\providecommand \href@noop [0]{\@secondoftwo}%
\providecommand \href [0]{\begingroup \@sanitize@url \@href}%
\providecommand \@href[1]{\@@startlink{#1}\@@href}%
\providecommand \@@href[1]{\endgroup#1\@@endlink}%
\providecommand \@sanitize@url [0]{\catcode `\\12\catcode `\$12\catcode
  `\&12\catcode `\#12\catcode `\^12\catcode `\_12\catcode `\%12\relax}%
\providecommand \@@startlink[1]{}%
\providecommand \@@endlink[0]{}%
\providecommand \url  [0]{\begingroup\@sanitize@url \@url }%
\providecommand \@url [1]{\endgroup\@href {#1}{\urlprefix }}%
\providecommand \urlprefix  [0]{URL }%
\providecommand \Eprint [0]{\href }%
\providecommand \doibase [0]{http://dx.doi.org/}%
\providecommand \selectlanguage [0]{\@gobble}%
\providecommand \bibinfo  [0]{\@secondoftwo}%
\providecommand \bibfield  [0]{\@secondoftwo}%
\providecommand \translation [1]{[#1]}%
\providecommand \BibitemOpen [0]{}%
\providecommand \bibitemStop [0]{}%
\providecommand \bibitemNoStop [0]{.\EOS\space}%
\providecommand \EOS [0]{\spacefactor3000\relax}%
\providecommand \BibitemShut  [1]{\csname bibitem#1\endcsname}%
\let\auto@bib@innerbib\@empty
\bibitem [{\citenamefont {Konig}\ \emph {et~al.}(2007)\citenamefont {Konig},
  \citenamefont {Wiedmann}, \citenamefont {Brune}, \citenamefont {Roth},
  \citenamefont {Buhmann}, \citenamefont {Molenkamp}, \citenamefont {Qi},\ and\
  \citenamefont {Zhang}}]{Konig2007}%
  \BibitemOpen
  \bibfield  {author} {\bibinfo {author} {\bibfnamefont {M.}~\bibnamefont
  {Konig}}, \bibinfo {author} {\bibfnamefont {S.}~\bibnamefont {Wiedmann}},
  \bibinfo {author} {\bibfnamefont {C.}~\bibnamefont {Brune}}, \bibinfo
  {author} {\bibfnamefont {A.}~\bibnamefont {Roth}}, \bibinfo {author}
  {\bibfnamefont {H.}~\bibnamefont {Buhmann}}, \bibinfo {author} {\bibfnamefont
  {L.~W.}\ \bibnamefont {Molenkamp}}, \bibinfo {author} {\bibfnamefont {X.-L.}\
  \bibnamefont {Qi}}, \ and\ \bibinfo {author} {\bibfnamefont {S.-C.}\
  \bibnamefont {Zhang}},\ }\href {\doibase 10.1126/science.1148047} {\bibfield
  {journal} {\bibinfo  {journal} {Science}\ }\textbf {\bibinfo {volume}
  {318}},\ \bibinfo {pages} {766} (\bibinfo {year} {2007})}\BibitemShut
  {NoStop}%
\bibitem [{\citenamefont {Xia}\ \emph {et~al.}(2009)\citenamefont {Xia},
  \citenamefont {Qian}, \citenamefont {Hsieh}, \citenamefont {Wray},
  \citenamefont {Pal}, \citenamefont {Lin}, \citenamefont {Bansil},
  \citenamefont {Grauer}, \citenamefont {Hor}, \citenamefont {Cava},\ and\
  \citenamefont {Hasan}}]{Xia2009}%
  \BibitemOpen
  \bibfield  {author} {\bibinfo {author} {\bibfnamefont {Y.}~\bibnamefont
  {Xia}}, \bibinfo {author} {\bibfnamefont {D.}~\bibnamefont {Qian}}, \bibinfo
  {author} {\bibfnamefont {D.}~\bibnamefont {Hsieh}}, \bibinfo {author}
  {\bibfnamefont {L.}~\bibnamefont {Wray}}, \bibinfo {author} {\bibfnamefont
  {A.}~\bibnamefont {Pal}}, \bibinfo {author} {\bibfnamefont {H.}~\bibnamefont
  {Lin}}, \bibinfo {author} {\bibfnamefont {A.}~\bibnamefont {Bansil}},
  \bibinfo {author} {\bibfnamefont {D.}~\bibnamefont {Grauer}}, \bibinfo
  {author} {\bibfnamefont {Y.~S.}\ \bibnamefont {Hor}}, \bibinfo {author}
  {\bibfnamefont {R.~J.}\ \bibnamefont {Cava}}, \ and\ \bibinfo {author}
  {\bibfnamefont {M.~Z.}\ \bibnamefont {Hasan}},\ }\href {\doibase
  10.1038/nphys1274} {\bibfield  {journal} {\bibinfo  {journal} {Nat. Phys.}\
  }\textbf {\bibinfo {volume} {5}},\ \bibinfo {pages} {398} (\bibinfo {year}
  {2009})}\BibitemShut {NoStop}%
\bibitem [{\citenamefont {Chen}\ \emph {et~al.}(2009)\citenamefont {Chen} \emph
  {et~al.}}]{Topological2009}%
  \BibitemOpen
  \bibfield  {author} {\bibinfo {author} {\bibfnamefont {Y.~L.}\ \bibnamefont
  {Chen}} \emph {et~al.},\ }\href {\doibase 10.1126/science.1173034} {\bibfield
   {journal} {\bibinfo  {journal} {Science}\ }\textbf {\bibinfo {volume}
  {325}},\ \bibinfo {pages} {178} (\bibinfo {year} {2009})}\BibitemShut
  {NoStop}%
\bibitem [{\citenamefont {Liu}\ \emph {et~al.}(2014{\natexlab{a}})\citenamefont
  {Liu} \emph {et~al.}}]{Liu2014}%
  \BibitemOpen
  \bibfield  {author} {\bibinfo {author} {\bibfnamefont {Z.~K.}\ \bibnamefont
  {Liu}} \emph {et~al.},\ }\href {\doibase 10.1126/science.1245085} {\bibfield
  {journal} {\bibinfo  {journal} {Science}\ }\textbf {\bibinfo {volume}
  {343}},\ \bibinfo {pages} {864} (\bibinfo {year}
  {2014}{\natexlab{a}})}\BibitemShut {NoStop}%
\bibitem [{\citenamefont {Liu}\ \emph {et~al.}(2014{\natexlab{b}})\citenamefont
  {Liu} \emph {et~al.}}]{Liu2014a}%
  \BibitemOpen
  \bibfield  {author} {\bibinfo {author} {\bibfnamefont {Z.~K.}\ \bibnamefont
  {Liu}} \emph {et~al.},\ }\href {http://www.ncbi.nlm.nih.gov/pubmed/24859642
  http://www.nature.com/doifinder/10.1038/nmat3990} {\bibfield  {journal}
  {\bibinfo  {journal} {Nat. Mater.}\ }\textbf {\bibinfo {volume} {13}},\
  \bibinfo {pages} {677} (\bibinfo {year} {2014}{\natexlab{b}})}\BibitemShut
  {NoStop}%
\bibitem [{\citenamefont {Neupane}\ \emph {et~al.}(2014)\citenamefont {Neupane}
  \emph {et~al.}}]{Neupane2014}%
  \BibitemOpen
  \bibfield  {author} {\bibinfo {author} {\bibfnamefont {M.}~\bibnamefont
  {Neupane}} \emph {et~al.},\ }\href
  {http://www.nature.com/doifinder/10.1038/ncomms4786} {\bibfield  {journal}
  {\bibinfo  {journal} {Nat. Commun.}\ }\textbf {\bibinfo {volume} {5}},\
  \bibinfo {pages} {3786} (\bibinfo {year} {2014})}\BibitemShut {NoStop}%
\bibitem [{\citenamefont {Xu}\ \emph {et~al.}(2015)\citenamefont {Xu} \emph
  {et~al.}}]{Xu2015}%
  \BibitemOpen
  \bibfield  {author} {\bibinfo {author} {\bibfnamefont {S.-Y.}\ \bibnamefont
  {Xu}} \emph {et~al.},\ }\href {\doibase 10.1126/science.aaa9297} {\bibfield
  {journal} {\bibinfo  {journal} {Science}\ }\textbf {\bibinfo {volume}
  {349}},\ \bibinfo {pages} {613} (\bibinfo {year} {2015})}\BibitemShut
  {NoStop}%
\bibitem [{\citenamefont {Lv}\ \emph {et~al.}(2015)\citenamefont {Lv} \emph
  {et~al.}}]{Lv2015}%
  \BibitemOpen
  \bibfield  {author} {\bibinfo {author} {\bibfnamefont {B.~Q.}\ \bibnamefont
  {Lv}} \emph {et~al.},\ }\href {\doibase 10.1103/PhysRevX.5.031013} {\bibfield
   {journal} {\bibinfo  {journal} {Phys. Rev. X}\ }\textbf {\bibinfo {volume}
  {5}},\ \bibinfo {pages} {031013} (\bibinfo {year} {2015})}\BibitemShut
  {NoStop}%
\bibitem [{\citenamefont {Jones}\ \emph {et~al.}(1982)\citenamefont {Jones},
  \citenamefont {Fuller}, \citenamefont {Wieting},\ and\ \citenamefont
  {Levy}}]{Jones1982}%
  \BibitemOpen
  \bibfield  {author} {\bibinfo {author} {\bibfnamefont {T.}~\bibnamefont
  {Jones}}, \bibinfo {author} {\bibfnamefont {W.}~\bibnamefont {Fuller}},
  \bibinfo {author} {\bibfnamefont {T.}~\bibnamefont {Wieting}}, \ and\
  \bibinfo {author} {\bibfnamefont {F.}~\bibnamefont {Levy}},\ }\href {\doibase
  10.1016/0038-1098(82)90008-4} {\bibfield  {journal} {\bibinfo  {journal}
  {Solid State Commun.}\ }\textbf {\bibinfo {volume} {42}},\ \bibinfo {pages}
  {793} (\bibinfo {year} {1982})}\BibitemShut {NoStop}%
\bibitem [{\citenamefont {Skelton}\ \emph {et~al.}(1982)\citenamefont
  {Skelton}, \citenamefont {Wieting}, \citenamefont {Wolf}, \citenamefont
  {Fuller}, \citenamefont {Gubser}, \citenamefont {Francavilla},\ and\
  \citenamefont {Levy}}]{Skelton1982}%
  \BibitemOpen
  \bibfield  {author} {\bibinfo {author} {\bibfnamefont {E.~F.}\ \bibnamefont
  {Skelton}}, \bibinfo {author} {\bibfnamefont {T.~J.}\ \bibnamefont
  {Wieting}}, \bibinfo {author} {\bibfnamefont {S.~A.}\ \bibnamefont {Wolf}},
  \bibinfo {author} {\bibfnamefont {W.~W.}\ \bibnamefont {Fuller}}, \bibinfo
  {author} {\bibfnamefont {D.~U.}\ \bibnamefont {Gubser}}, \bibinfo {author}
  {\bibfnamefont {T.~L.}\ \bibnamefont {Francavilla}}, \ and\ \bibinfo {author}
  {\bibfnamefont {F.}~\bibnamefont {Levy}},\ }\href {\doibase
  10.1016/0038-1098(82)91016-X} {\bibfield  {journal} {\bibinfo  {journal}
  {Solid State Commun.}\ }\textbf {\bibinfo {volume} {42}},\ \bibinfo {pages}
  {1} (\bibinfo {year} {1982})}\BibitemShut {NoStop}%
\bibitem [{\citenamefont {Tritt}\ \emph {et~al.}(1999)\citenamefont {Tritt},
  \citenamefont {Lowhorn}, \citenamefont {Littleton}, \citenamefont {Pope},
  \citenamefont {Feger},\ and\ \citenamefont {Kolis}}]{Tritt1999}%
  \BibitemOpen
  \bibfield  {author} {\bibinfo {author} {\bibfnamefont {T.~M.}\ \bibnamefont
  {Tritt}}, \bibinfo {author} {\bibfnamefont {N.~D.}\ \bibnamefont {Lowhorn}},
  \bibinfo {author} {\bibfnamefont {R.~T.}\ \bibnamefont {Littleton}}, \bibinfo
  {author} {\bibfnamefont {A.}~\bibnamefont {Pope}}, \bibinfo {author}
  {\bibfnamefont {C.~R.}\ \bibnamefont {Feger}}, \ and\ \bibinfo {author}
  {\bibfnamefont {J.~W.}\ \bibnamefont {Kolis}},\ }\href {\doibase
  10.1103/PhysRevB.60.7816} {\bibfield  {journal} {\bibinfo  {journal} {Phys.
  Rev. B}\ }\textbf {\bibinfo {volume} {60}},\ \bibinfo {pages} {7816}
  (\bibinfo {year} {1999})}\BibitemShut {NoStop}%
\bibitem [{\citenamefont {McIlroy}\ \emph {et~al.}(2004)\citenamefont {McIlroy}
  \emph {et~al.}}]{McIlroy2004}%
  \BibitemOpen
  \bibfield  {author} {\bibinfo {author} {\bibfnamefont {D.~N.}\ \bibnamefont
  {McIlroy}} \emph {et~al.},\ }\href {\doibase 10.1088/0953-8984/16/30/L02}
  {\bibfield  {journal} {\bibinfo  {journal} {J. Phys. Condens. Matter}\
  }\textbf {\bibinfo {volume} {16}},\ \bibinfo {pages} {L359} (\bibinfo {year}
  {2004})}\BibitemShut {NoStop}%
\bibitem [{\citenamefont {Zhou}\ \emph {et~al.}(2016)\citenamefont {Zhou} \emph
  {et~al.}}]{Zhou2016}%
  \BibitemOpen
  \bibfield  {author} {\bibinfo {author} {\bibfnamefont {Y.}~\bibnamefont
  {Zhou}} \emph {et~al.},\ }\href {\doibase 10.1073/pnas.1601262113} {\bibfield
   {journal} {\bibinfo  {journal} {Proc. Natl. Acad. Sci.}\ }\textbf {\bibinfo
  {volume} {113}},\ \bibinfo {pages} {2904} (\bibinfo {year}
  {2016})}\BibitemShut {NoStop}%
\bibitem [{\citenamefont {Weng}\ \emph {et~al.}(2014)\citenamefont {Weng},
  \citenamefont {Dai},\ and\ \citenamefont {Fang}}]{Weng2014}%
  \BibitemOpen
  \bibfield  {author} {\bibinfo {author} {\bibfnamefont {H.}~\bibnamefont
  {Weng}}, \bibinfo {author} {\bibfnamefont {X.}~\bibnamefont {Dai}}, \ and\
  \bibinfo {author} {\bibfnamefont {Z.}~\bibnamefont {Fang}},\ }\href {\doibase
  10.1103/PhysRevX.4.011002} {\bibfield  {journal} {\bibinfo  {journal} {Phys.
  Rev. X}\ }\textbf {\bibinfo {volume} {4}},\ \bibinfo {pages} {011002}
  (\bibinfo {year} {2014})}\BibitemShut {NoStop}%
\bibitem [{\citenamefont {Manzoni}\ \emph
  {et~al.}(2016{\natexlab{a}})\citenamefont {Manzoni} \emph
  {et~al.}}]{Manzoni2016}%
  \BibitemOpen
  \bibfield  {author} {\bibinfo {author} {\bibfnamefont {G.}~\bibnamefont
  {Manzoni}} \emph {et~al.},\ }\href {\doibase 10.1103/PhysRevLett.117.237601}
  {\bibfield  {journal} {\bibinfo  {journal} {Phys. Rev. Lett.}\ }\textbf
  {\bibinfo {volume} {117}},\ \bibinfo {pages} {237601} (\bibinfo {year}
  {2016}{\natexlab{a}})}\BibitemShut {NoStop}%
\bibitem [{\citenamefont {Li}\ \emph {et~al.}(2016{\natexlab{a}})\citenamefont
  {Li} \emph {et~al.}}]{Li2016}%
  \BibitemOpen
  \bibfield  {author} {\bibinfo {author} {\bibfnamefont {Q.}~\bibnamefont {Li}}
  \emph {et~al.},\ }\href {\doibase 10.1038/nphys3648} {\bibfield  {journal}
  {\bibinfo  {journal} {Nat. Phys.}\ }\textbf {\bibinfo {volume} {12}},\
  \bibinfo {pages} {550} (\bibinfo {year} {2016}{\natexlab{a}})}\BibitemShut
  {NoStop}%
\bibitem [{\citenamefont {Shen}\ \emph {et~al.}(2016)\citenamefont {Shen} \emph
  {et~al.}}]{Shen2016}%
  \BibitemOpen
  \bibfield  {author} {\bibinfo {author} {\bibfnamefont {L.}~\bibnamefont
  {Shen}} \emph {et~al.},\ }\href {\doibase 10.1016/j.elspec.2016.10.007}
  {\bibfield  {journal} {\bibinfo  {journal} {J. Electron Spectros. Relat.
  Phenom.}\ } (\bibinfo {year} {2016}),\
  10.1016/j.elspec.2016.10.007}\BibitemShut {NoStop}%
\bibitem [{\citenamefont {Chen}\ \emph
  {et~al.}(2015{\natexlab{a}})\citenamefont {Chen}, \citenamefont {Chen},
  \citenamefont {Song}, \citenamefont {Schneeloch}, \citenamefont {Gu},
  \citenamefont {Wang},\ and\ \citenamefont {Wang}}]{Chen2015}%
  \BibitemOpen
  \bibfield  {author} {\bibinfo {author} {\bibfnamefont {R.~Y.}\ \bibnamefont
  {Chen}}, \bibinfo {author} {\bibfnamefont {Z.~G.}\ \bibnamefont {Chen}},
  \bibinfo {author} {\bibfnamefont {X.~Y.}\ \bibnamefont {Song}}, \bibinfo
  {author} {\bibfnamefont {J.~A.}\ \bibnamefont {Schneeloch}}, \bibinfo
  {author} {\bibfnamefont {G.~D.}\ \bibnamefont {Gu}}, \bibinfo {author}
  {\bibfnamefont {F.}~\bibnamefont {Wang}}, \ and\ \bibinfo {author}
  {\bibfnamefont {N.~L.}\ \bibnamefont {Wang}},\ }\href {\doibase
  10.1103/PhysRevLett.115.176404} {\bibfield  {journal} {\bibinfo  {journal}
  {Phys. Rev. Lett.}\ }\textbf {\bibinfo {volume} {115}},\ \bibinfo {pages}
  {176404} (\bibinfo {year} {2015}{\natexlab{a}})}\BibitemShut {NoStop}%
\bibitem [{\citenamefont {Chen}\ \emph
  {et~al.}(2015{\natexlab{b}})\citenamefont {Chen}, \citenamefont {Zhang},
  \citenamefont {Schneeloch}, \citenamefont {Zhang}, \citenamefont {Li},
  \citenamefont {Gu},\ and\ \citenamefont {Wang}}]{Chen2015a}%
  \BibitemOpen
  \bibfield  {author} {\bibinfo {author} {\bibfnamefont {R.~Y.}\ \bibnamefont
  {Chen}}, \bibinfo {author} {\bibfnamefont {S.~J.}\ \bibnamefont {Zhang}},
  \bibinfo {author} {\bibfnamefont {J.~A.}\ \bibnamefont {Schneeloch}},
  \bibinfo {author} {\bibfnamefont {C.}~\bibnamefont {Zhang}}, \bibinfo
  {author} {\bibfnamefont {Q.}~\bibnamefont {Li}}, \bibinfo {author}
  {\bibfnamefont {G.~D.}\ \bibnamefont {Gu}}, \ and\ \bibinfo {author}
  {\bibfnamefont {N.~L.}\ \bibnamefont {Wang}},\ }\href {\doibase
  10.1103/PhysRevB.92.075107} {\bibfield  {journal} {\bibinfo  {journal} {Phys.
  Rev. B}\ }\textbf {\bibinfo {volume} {92}},\ \bibinfo {pages} {075107}
  (\bibinfo {year} {2015}{\natexlab{b}})}\BibitemShut {NoStop}%
\bibitem [{\citenamefont {Zheng}\ \emph {et~al.}(2016)\citenamefont {Zheng}
  \emph {et~al.}}]{Zheng2016}%
  \BibitemOpen
  \bibfield  {author} {\bibinfo {author} {\bibfnamefont {G.}~\bibnamefont
  {Zheng}} \emph {et~al.},\ }\href {\doibase 10.1103/PhysRevB.93.115414}
  {\bibfield  {journal} {\bibinfo  {journal} {Phys. Rev. B}\ }\textbf {\bibinfo
  {volume} {93}},\ \bibinfo {pages} {115414} (\bibinfo {year}
  {2016})}\BibitemShut {NoStop}%
\bibitem [{\citenamefont {Wu}\ \emph {et~al.}(2016)\citenamefont {Wu} \emph
  {et~al.}}]{Wu2016}%
  \BibitemOpen
  \bibfield  {author} {\bibinfo {author} {\bibfnamefont {R.}~\bibnamefont {Wu}}
  \emph {et~al.},\ }\href {\doibase 10.1103/PhysRevX.6.021017} {\bibfield
  {journal} {\bibinfo  {journal} {Phys. Rev. X}\ }\textbf {\bibinfo {volume}
  {6}},\ \bibinfo {pages} {021017} (\bibinfo {year} {2016})}\BibitemShut
  {NoStop}%
\bibitem [{\citenamefont {Li}\ \emph {et~al.}(2016{\natexlab{b}})\citenamefont
  {Li} \emph {et~al.}}]{Li2016a}%
  \BibitemOpen
  \bibfield  {author} {\bibinfo {author} {\bibfnamefont {X.-B.}\ \bibnamefont
  {Li}} \emph {et~al.},\ }\href {\doibase 10.1103/PhysRevLett.116.176803}
  {\bibfield  {journal} {\bibinfo  {journal} {Phys. Rev. Lett.}\ }\textbf
  {\bibinfo {volume} {116}},\ \bibinfo {pages} {176803} (\bibinfo {year}
  {2016}{\natexlab{b}})}\BibitemShut {NoStop}%
\bibitem [{\citenamefont {Manzoni}\ \emph {et~al.}(2015)\citenamefont {Manzoni}
  \emph {et~al.}}]{Manzoni2015}%
  \BibitemOpen
  \bibfield  {author} {\bibinfo {author} {\bibfnamefont {G.}~\bibnamefont
  {Manzoni}} \emph {et~al.},\ }\href {\doibase 10.1103/PhysRevLett.115.207402}
  {\bibfield  {journal} {\bibinfo  {journal} {Phys. Rev. Lett.}\ }\textbf
  {\bibinfo {volume} {115}},\ \bibinfo {pages} {207402} (\bibinfo {year}
  {2015})}\BibitemShut {NoStop}%
\bibitem [{\citenamefont {Zhang}\ \emph {et~al.}()\citenamefont {Zhang} \emph
  {et~al.}}]{Zhang2016}%
  \BibitemOpen
  \bibfield  {author} {\bibinfo {author} {\bibfnamefont {Y.}~\bibnamefont
  {Zhang}} \emph {et~al.},\ }\href {http://arxiv.org/abs/1602.03576} {\
  }\Eprint {http://arxiv.org/abs/1602.03576} {arXiv:1602.03576} \BibitemShut
  {NoStop}%
\bibitem [{\citenamefont {Manzoni}\ \emph
  {et~al.}(2016{\natexlab{b}})\citenamefont {Manzoni} \emph
  {et~al.}}]{Manzoni2016a}%
  \BibitemOpen
  \bibfield  {author} {\bibinfo {author} {\bibfnamefont {G.}~\bibnamefont
  {Manzoni}} \emph {et~al.},\ }\href {\doibase 10.1016/j.elspec.2016.09.006}
  {\bibfield  {journal} {\bibinfo  {journal} {J. Electron Spectros. Relat.
  Phenom.}\ } (\bibinfo {year} {2016}{\natexlab{b}}),\
  10.1016/j.elspec.2016.09.006}\BibitemShut {NoStop}%
\bibitem [{\citenamefont {Moreschini}\ \emph {et~al.}(2016)\citenamefont
  {Moreschini} \emph {et~al.}}]{Moreschini2016}%
  \BibitemOpen
  \bibfield  {author} {\bibinfo {author} {\bibfnamefont {L.}~\bibnamefont
  {Moreschini}} \emph {et~al.},\ }\href {\doibase 10.1103/PhysRevB.94.081101}
  {\bibfield  {journal} {\bibinfo  {journal} {Phys. Rev. B}\ }\textbf {\bibinfo
  {volume} {94}},\ \bibinfo {pages} {081101} (\bibinfo {year}
  {2016})}\BibitemShut {NoStop}%
\bibitem [{\citenamefont {Lv}\ \emph {et~al.}(2016)\citenamefont {Lv} \emph
  {et~al.}}]{Lv2016}%
  \BibitemOpen
  \bibfield  {author} {\bibinfo {author} {\bibfnamefont {Y.-Y.}\ \bibnamefont
  {Lv}} \emph {et~al.},\ }\href {\doibase 10.1016/j.jcrysgro.2016.04.042}
  {\bibfield  {journal} {\bibinfo  {journal} {J. Cryst. Growth}\ }\textbf
  {\bibinfo {volume} {457}},\ \bibinfo {pages} {250} (\bibinfo {year}
  {2016})}\BibitemShut {NoStop}%
\bibitem [{\citenamefont {Pan}\ \emph {et~al.}(2012)\citenamefont {Pan},
  \citenamefont {Fedorov}, \citenamefont {Gardner}, \citenamefont {Lee},
  \citenamefont {Chu},\ and\ \citenamefont {Valla}}]{Pan2012}%
  \BibitemOpen
  \bibfield  {author} {\bibinfo {author} {\bibfnamefont {Z.-H.}\ \bibnamefont
  {Pan}}, \bibinfo {author} {\bibfnamefont {A.~V.}\ \bibnamefont {Fedorov}},
  \bibinfo {author} {\bibfnamefont {D.}~\bibnamefont {Gardner}}, \bibinfo
  {author} {\bibfnamefont {Y.~S.}\ \bibnamefont {Lee}}, \bibinfo {author}
  {\bibfnamefont {S.}~\bibnamefont {Chu}}, \ and\ \bibinfo {author}
  {\bibfnamefont {T.}~\bibnamefont {Valla}},\ }\href {\doibase
  10.1103/PhysRevLett.108.187001} {\bibfield  {journal} {\bibinfo  {journal}
  {Phys. Rev. Lett.}\ }\textbf {\bibinfo {volume} {108}},\ \bibinfo {pages}
  {187001} (\bibinfo {year} {2012})}\BibitemShut {NoStop}%
\bibitem [{\citenamefont {Kondo}\ \emph {et~al.}(2013)\citenamefont {Kondo}
  \emph {et~al.}}]{Kondo2013}%
  \BibitemOpen
  \bibfield  {author} {\bibinfo {author} {\bibfnamefont {T.}~\bibnamefont
  {Kondo}} \emph {et~al.},\ }\href {\doibase 10.1103/PhysRevLett.110.217601}
  {\bibfield  {journal} {\bibinfo  {journal} {Phys. Rev. Lett.}\ }\textbf
  {\bibinfo {volume} {110}},\ \bibinfo {pages} {217601} (\bibinfo {year}
  {2013})}\BibitemShut {NoStop}%
\bibitem [{\citenamefont {He}\ \emph {et~al.}()\citenamefont {He},
  \citenamefont {Wang},\ and\ \citenamefont {Shen}}]{He2016}%
  \BibitemOpen
  \bibfield  {author} {\bibinfo {author} {\bibfnamefont {Y.}~\bibnamefont
  {He}}, \bibinfo {author} {\bibfnamefont {Y.}~\bibnamefont {Wang}}, \ and\
  \bibinfo {author} {\bibfnamefont {Z.-X.}\ \bibnamefont {Shen}},\ }\href
  {http://arxiv.org/abs/1612.07880} {\ }\Eprint
  {http://arxiv.org/abs/1612.07880} {arXiv:1612.07880} \BibitemShut {NoStop}%
\bibitem [{\citenamefont {Lindroos}\ and\ \citenamefont
  {Bansil}(1996)}]{Lindroos1996}%
  \BibitemOpen
  \bibfield  {author} {\bibinfo {author} {\bibfnamefont {M.}~\bibnamefont
  {Lindroos}}\ and\ \bibinfo {author} {\bibfnamefont {A.}~\bibnamefont
  {Bansil}},\ }\href {\doibase 10.1103/PhysRevLett.77.2985} {\bibfield
  {journal} {\bibinfo  {journal} {Phys. Rev. Lett.}\ }\textbf {\bibinfo
  {volume} {77}},\ \bibinfo {pages} {2985} (\bibinfo {year}
  {1996})}\BibitemShut {NoStop}%
\bibitem [{\citenamefont {Miller}\ \emph {et~al.}(2015)\citenamefont {Miller}
  \emph {et~al.}}]{Miller2015}%
  \BibitemOpen
  \bibfield  {author} {\bibinfo {author} {\bibfnamefont {T.~L.}\ \bibnamefont
  {Miller}} \emph {et~al.},\ }\href {\doibase 10.1103/PhysRevB.91.085109}
  {\bibfield  {journal} {\bibinfo  {journal} {Phys. Rev. B}\ }\textbf {\bibinfo
  {volume} {91}},\ \bibinfo {pages} {085109} (\bibinfo {year}
  {2015})}\BibitemShut {NoStop}%
\bibitem [{\citenamefont {{\"{A}}rr{\"{a}}l{\"{a}}}\ \emph
  {et~al.}(2016)\citenamefont {{\"{A}}rr{\"{a}}l{\"{a}}} \emph
  {et~al.}}]{Arrala2016}%
  \BibitemOpen
  \bibfield  {author} {\bibinfo {author} {\bibfnamefont {M.}~\bibnamefont
  {{\"{A}}rr{\"{a}}l{\"{a}}}} \emph {et~al.},\ }\href {\doibase
  10.1103/PhysRevB.94.155144} {\bibfield  {journal} {\bibinfo  {journal} {Phys.
  Rev. B}\ }\textbf {\bibinfo {volume} {94}},\ \bibinfo {pages} {155144}
  (\bibinfo {year} {2016})}\BibitemShut {NoStop}%
\end{thebibliography}%

\end{document}